\documentclass[prl, aps, 12pt, showkeys, showpacs,tightenlines] {revtex4}
\usepackage{amsmath}
\usepackage{epsfig}
\usepackage{hyperref}
\begin{document}
\title{Entropy Production and Viscosity of A Dilute Gas}
\author{Yong-Jun Zhang}
\email{yong.j.zhang@gmail.com}
\affiliation{Science College, Liaoning Technical University, Fuxin, Liaoning 123000, China}

\begin{abstract}
It is known that the viscosity of a dilute gas can be derived by using kinetic theory. We present here a new derivation by using two entropy production principles: the steepest entropy ascent (SEA) principle and the maximum entropy production (MEP) principle. The known result is reproduced in a similar form. 
\end{abstract}
\keywords{entropy production; SEA principle; MEP principle; viscosity; dilute gas}
\pacs{51.20.+d, 51.30.+i, 47.15.Fe, 05.70.Ln}
\maketitle

\section{Introduction}
Viscosity (or shear viscosity), $\eta$, is a coefficient that appears in Newton's law of viscosity,
\begin{equation}\label{jv}
        \tau_{zx}=\eta \frac{dv_x}{dz},
\end{equation}
where $\tau_{zx}$ is the shear stress and $\frac{dv_x}{dz}$ is the velocity gradient. For a dilute gas, the viscosity can be derived by using kinetic theory \cite{gas1}, with the result
\begin{equation}\label{kd}
	\eta=\frac{1}{3}\rho \bar{v}\lambda,
\end{equation}
where $\rho$ is the density, $\bar{v}$ is the mean speed and $\lambda$ is the mean free path. Using kinetic theory to derive the viscosity is simple. We know that, for a given velocity gradient $\frac{dv_x}{dz}$, the $x$-component of the molecular momentum increases at rate $M\frac{dv_x}{dz}$ in the $z$ direction; here $M$ is the mass of a molecule. We also know that in a given direction, on average a molecule needs to travel $\lambda_z$ to make a collision. So one may study a layer of a dilute gas with thickness $2\lambda_z$ in the $z$ direction and view the dilute gas as two layers, each having height $\lambda_z$. Two such layers of dilute gas will exchange molecules and momentum. For each pair of molecules to be exchanged, the amount of $x$-momentum to be exchanged is twice $\lambda_zM\frac{dv_x}{dz}$. During unit time, across unit cross-sectional area, the number of molecules to be exchanged is $n\bar{v}_z$, where $n$ is the number density. The corresponding $x$-momentum to be exchanged is $n\bar{v}_z\lambda_zM\frac{dv_x}{dz}$, which is actually the shear stress $\tau_{zx}$. By using Eq. (\ref{jv}) and the relations $\lambda_z=\frac{1}{\sqrt{3}}\lambda$, $\bar{v}_z=\frac{1}{\sqrt{3}}\bar{v}$ and $\rho=nM$, the viscosity Eq. (\ref{kd}) is obtained.

We shall present in this letter a new derivation of the viscosity of a dilute gas, by analogy with a derivation of the thermal conductivity of a dilute gas \cite{thermalconductivity}. The new derivation will use the entropy production. Here, two aspects of the entropy production will be used: (1) for a given state, it evolves in the direction of the steepest entropy ascent; (2) among many candidates, the actual steady state has the maximum entropy production. We call the first the steepest entropy ascent (SEA) principle and the second the maximum entropy production (MEP) principle. The name of "SEA" was originally introduced by Beretta \cite{Beretta3,Beretta32,Beretta4,Beretta5,Beretta6} to postulate that a system always evolves to the direction having the steepest entropy ascent. The MEP principle has been used by Paltridge \cite{Paltridge1, Paltridge2, Paltridge3} to study the Earth's climate by postulating that the steady state of the atmosphere has the maximum entropy production. Though in their original proposed form, the SEA principle and the MEP principle overlap and one even can be alternative to the other \cite{Beretta6}, we use here their names to emphasize two separate aspects of the entropy production. There exist many studies of entropy production theory itself; see, for example \cite{Ziegler,MEPP,Dewar1,Dewar2,Dewar_book,Jaynes1,Jaynes2,Grinstein,Bruers,Niven,Croatica}. This paper provides an application.

In order to derive the viscosity of a dilute gas, we shall study a dilute gas confined between two parallel plates where the bottom plate is stationary and the top plate is subject to a constant shear stress. Let the dilute gas be subject to three restrictions: (1) the density $\rho$ or $n$ is constant, and so we can write the number of molecules as $N=nA\Delta z$, where $A$ is the cross-sectional area and $\Delta z$ is the height of the dilute gas; (2) $N$ is large, $N\gg 1$, and so we can apply some approximations; (3) the shear stress is small, and so only laminar flow appears, and the velocity gradient is small. Such a dilute gas will carry a steady velocity gradient from whose expression the viscosity can be extracted. The expression of the steady velocity gradient will be determined by using the two entropy production principles. To do that, we shall first obtain the entropy of a dilute gas with respect to the velocity gradient.

\section{Entropy of a Dilute Gas having height $\lambda_z$}
\begin{figure}[htbp]
  \begin{center}
    \mbox{\epsfxsize=5.0cm\epsfysize=5.0cm\epsffile{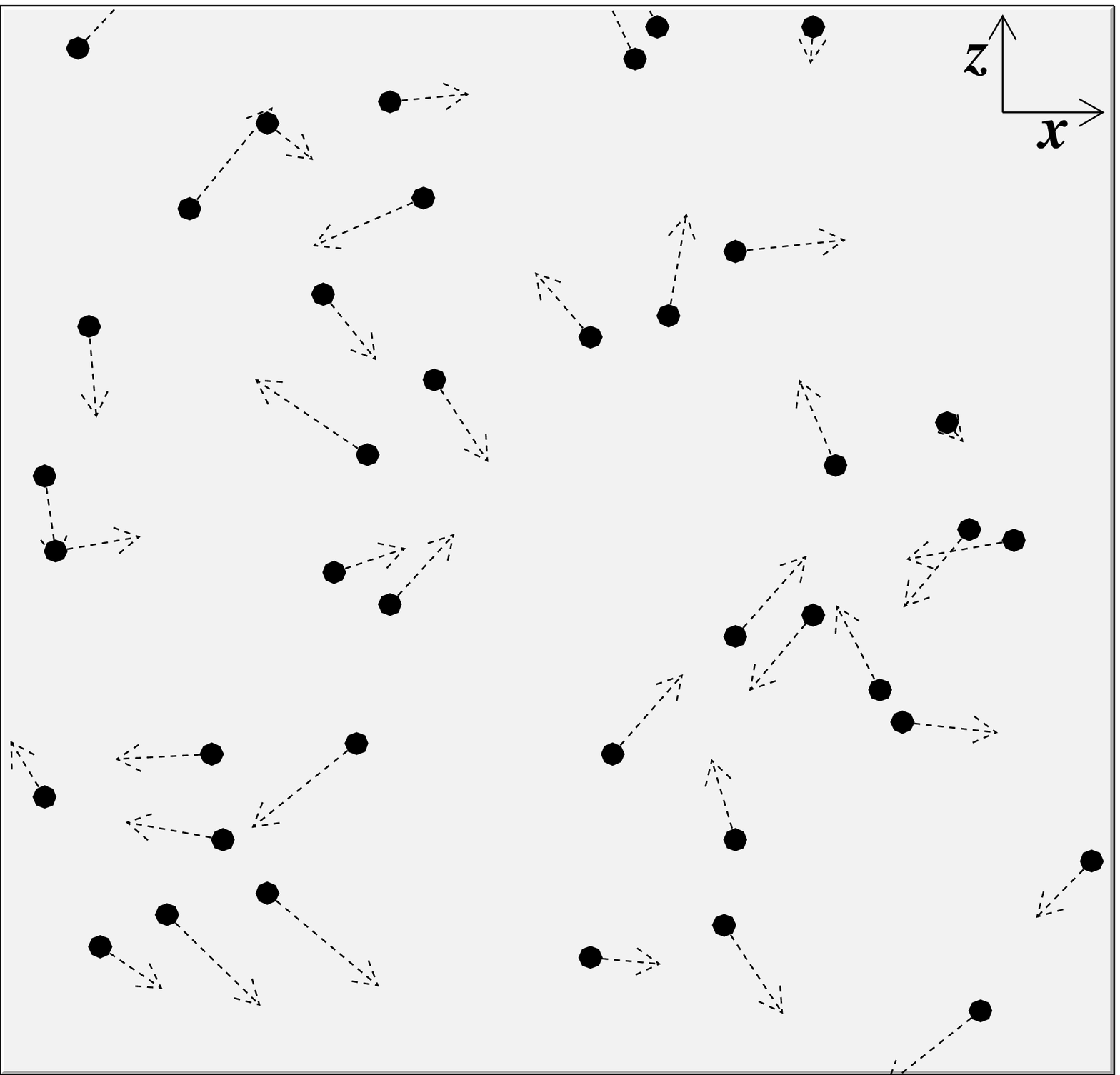}
	\ \epsfxsize=5.0cm\epsfysize=5.0cm\epsffile{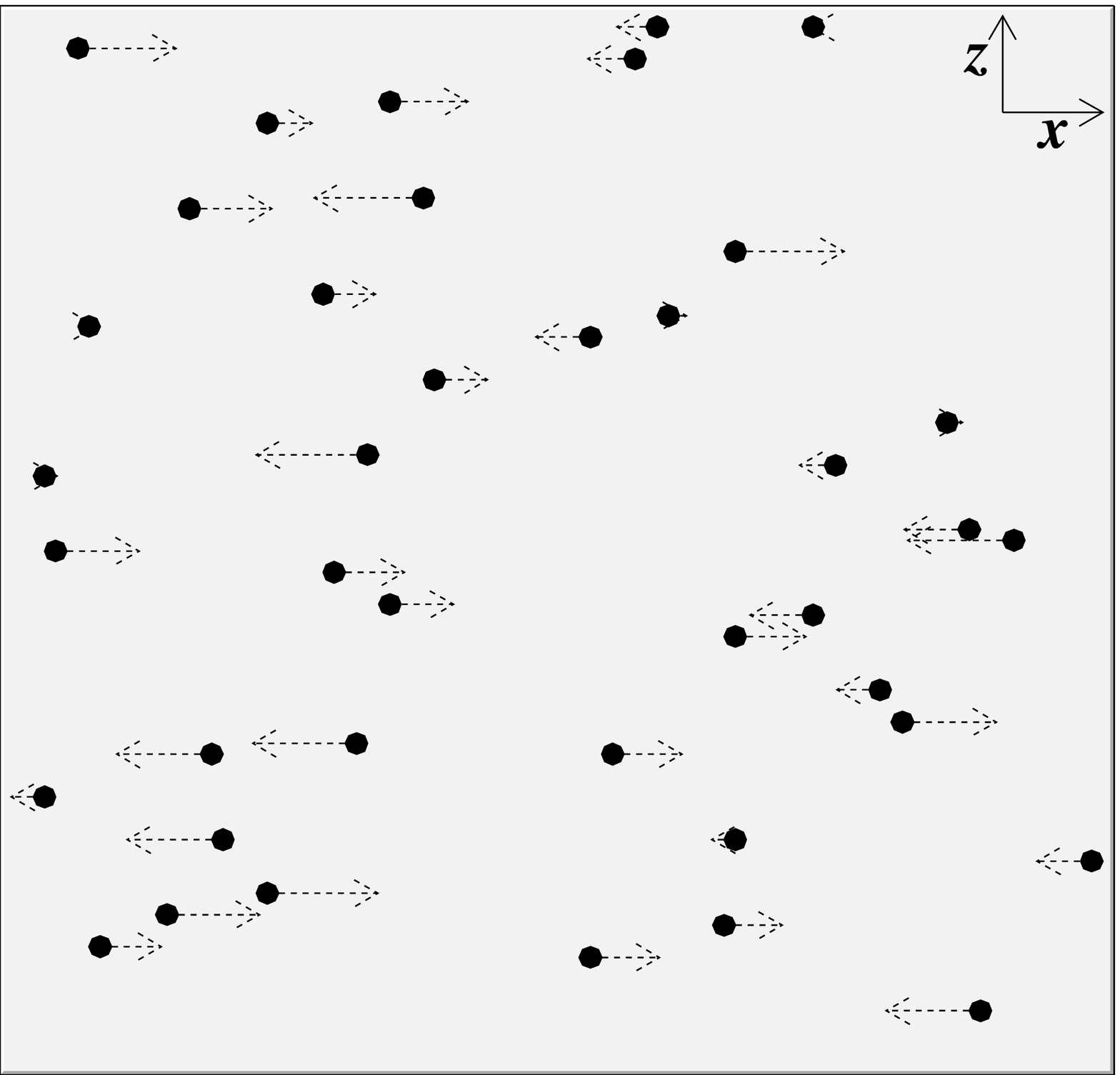}
	\ \epsfxsize=5.0cm\epsfysize=5.0cm\epsffile{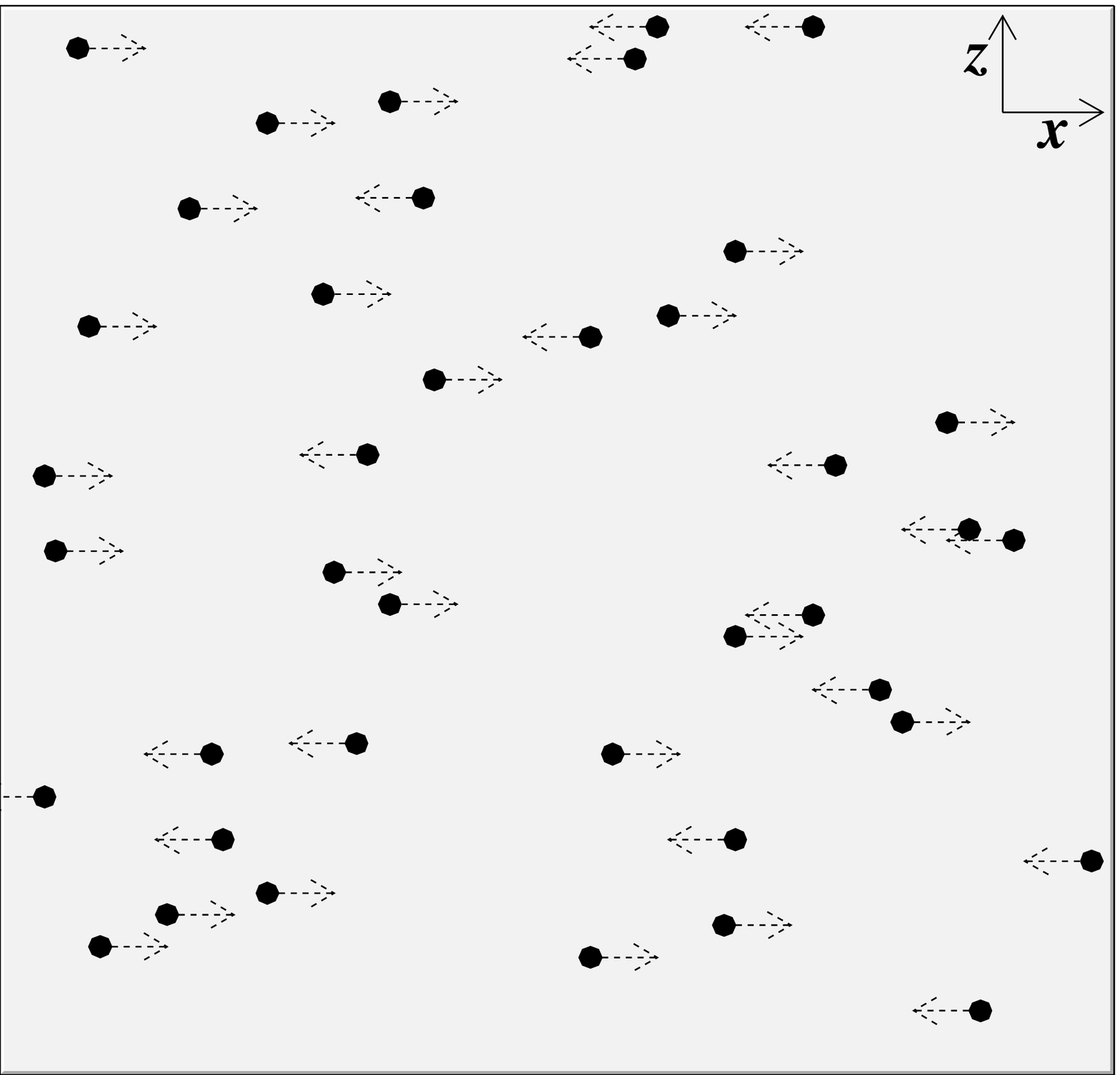}} 
  \end{center}
\caption{The molecular speed distribution of a dilute gas in the equilibrium state. The left graph shows only the molecular speed in the $x$ and $z$ direction. The middle graph shows only the molecular speed in the $x$ direction. The right graph shows only whether a molecule moves along the $+x$ direction or the $-x$ direction. The right graph can also be viewed as that a molecule moves either in the $+x$ direction or the $-x$ direction with the same speed $\bar{v}_z$, which satisfies $\bar{v}_x=\bar{v}_y=\bar{v}_z=\frac{\bar{v}}{\sqrt{3}}$.\label{equilibrium}}
\end{figure}

\begin{figure}[htbp]
  \begin{center}
	\mbox{\epsfxsize=5.0cm\epsfysize=5.0cm\epsffile{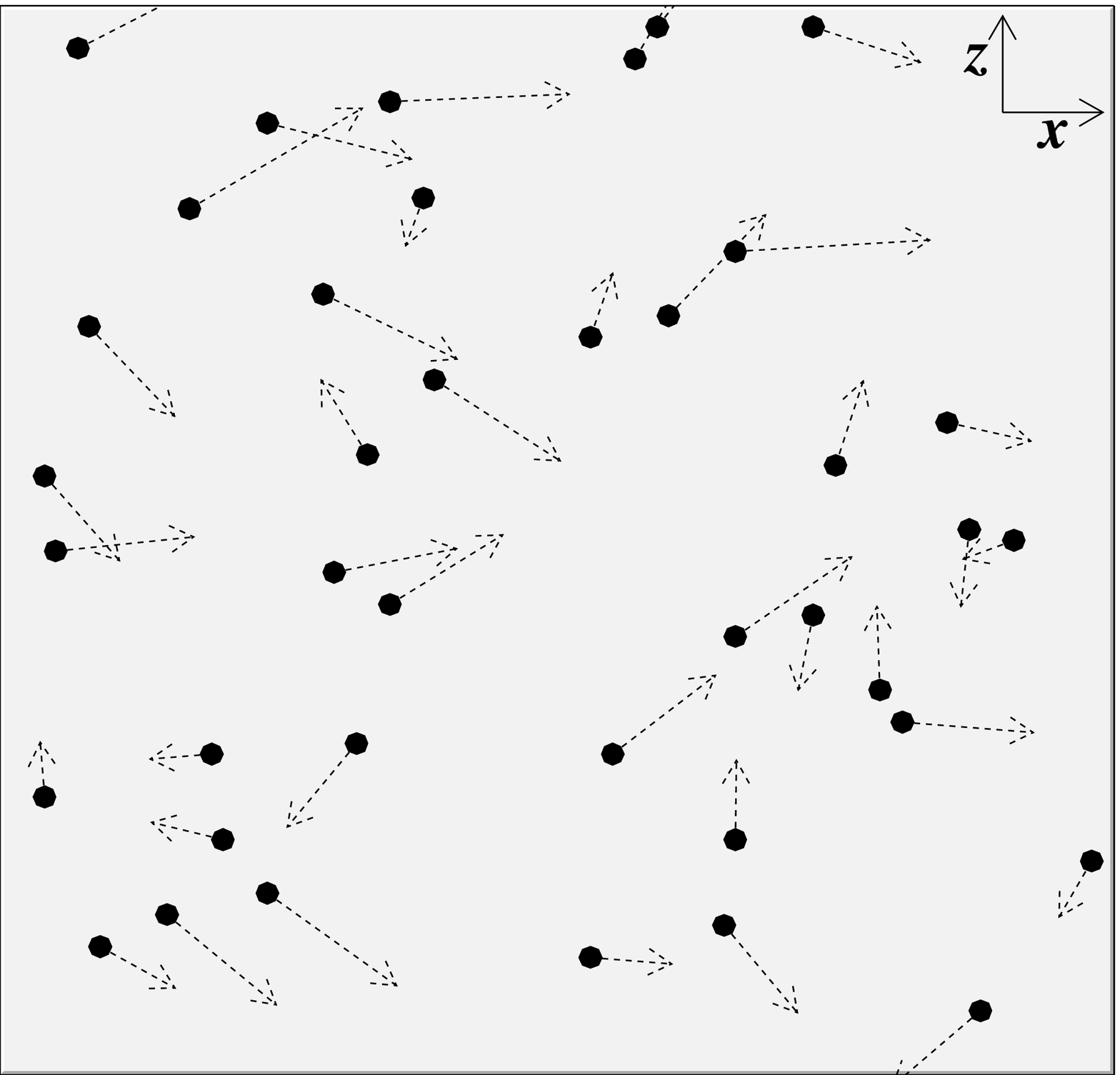}
	    \ \epsfxsize=5.0cm\epsfysize=5.0cm\epsffile{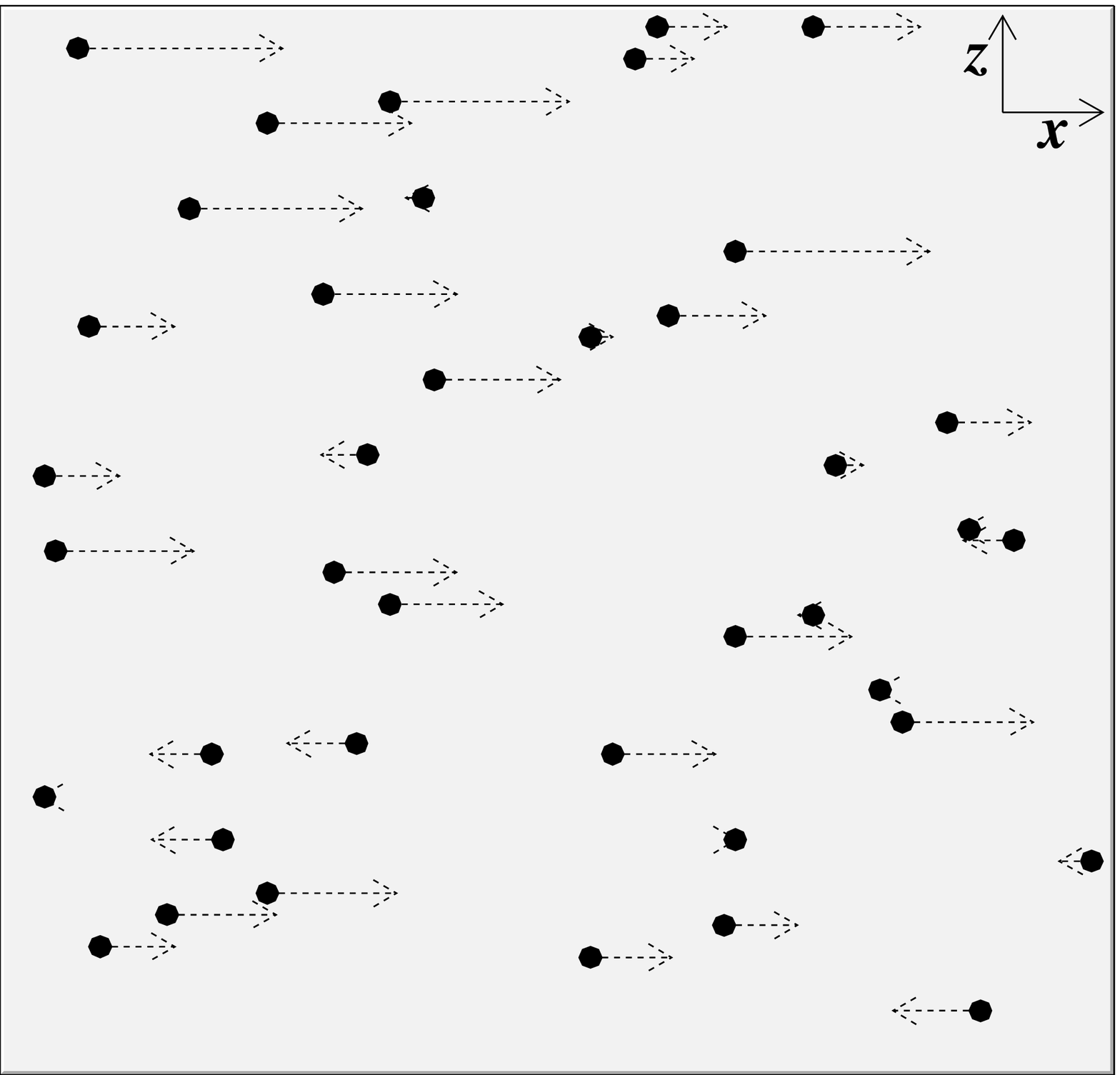}
	    \ \epsfxsize=5.0cm\epsfysize=5.0cm\epsffile{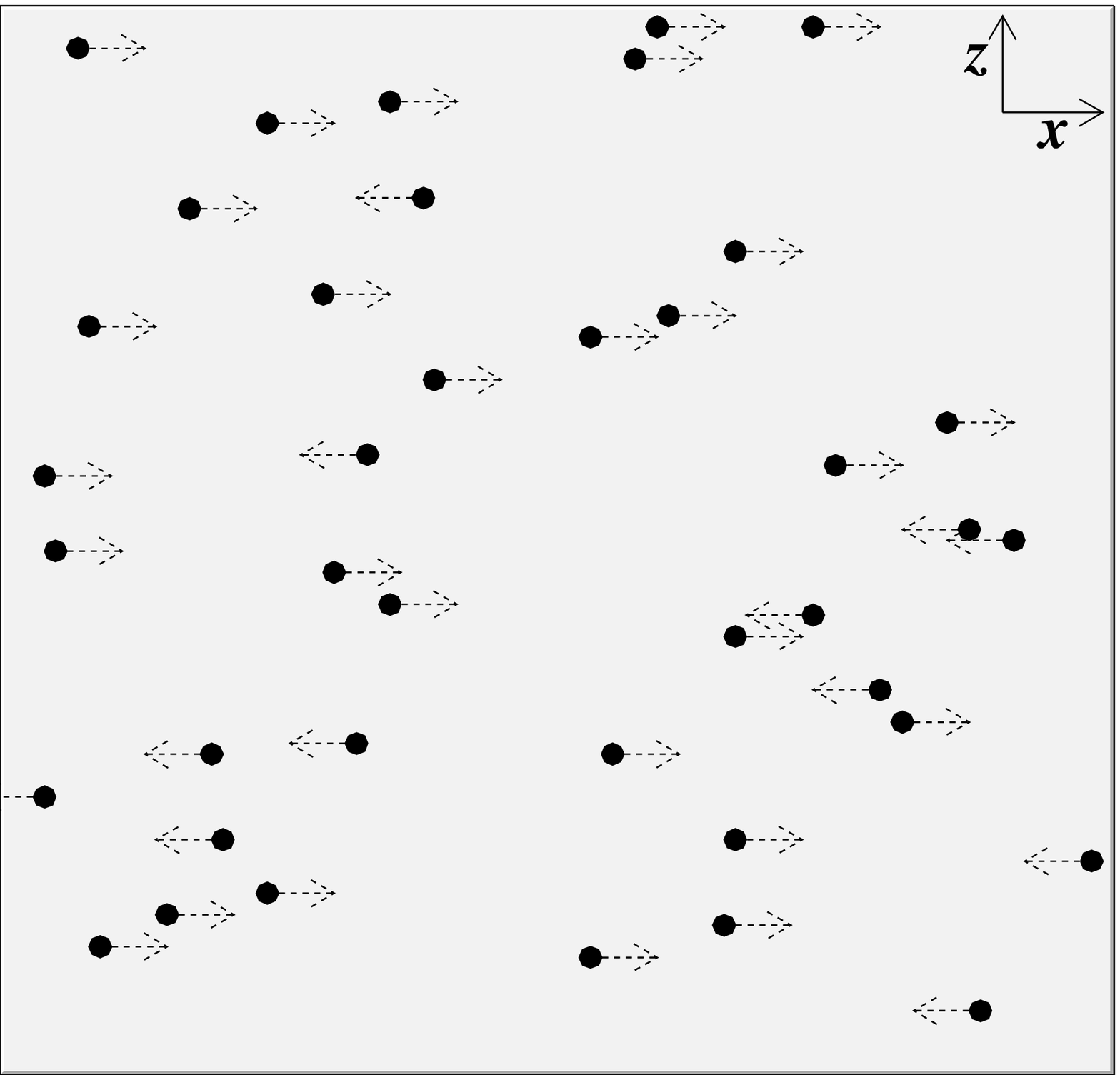}}
  \end{center}
\caption{The molecular speed distribution of a dilute gas carrying a velocity gradient $\frac{dv_x}{dz}$. The dilute gas is on a stationary plate, its height is $\lambda_z$, and it consists of $N$ molecules of which $k$ molecules move in the $+x$ direction. It can be viewed as a fluctuation of the equilibrium state in Fig. \ref{equilibrium} for which $k\approx \frac{N}{2}$, but we have here $k=\frac{N}{2}+\frac{N\lambda_z}{4\bar{v}_z}\frac{dv_x}{dz}$. For a given $k$, the number of corresponding microscopic states is $C_N^k$ and the entropy is $S(k)=k_B\ln C_N^k$.
\label{gradient}}
\end{figure}

In this section, we study a dilute gas having height $\lambda_z$ and estimate its entropy with respect to the velocity gradient. 
As Fig. \ref{equilibrium} and Fig. \ref{gradient} show, for each molecule, we only need to study how it moves in the $x$ direction, given that the velocity gradient is $\frac{dv_x}{dz}$. So we may simply consider that each molecule moves either in the $+x$ direction or the $-x$ direction at the same speed $\bar{v}_z$.  

Let us study a macroscopic state in which $k$ molecules out of $N$ molecules move in the $+x$ direction. The number of corresponding microscopic states is $\Omega(k)=C_N^k$, and by using Boltzmann's entropy formula the corresponding entropy is
\begin{equation} 
	S(k)=k_B\ln\Omega(k)=k_B\ln C_N^k=k_B\ln \frac{N!}{k!(N-k)!}
\end{equation}
Given the velocity gradient $\frac{dv_x}{dz}$ being small, $k$ is comparable to $\frac{N}{2}$, $k\sim \frac{N}{2}$. Given $N$ being large, Stirling's approximation can be used, $\ln N!\approx N\ln N-N$ and $\ln k!\approx k\ln k-k$. Then by using Taylor's series approximation $\ln (1+x)\approx x-\frac{x^2}{2}$, where $x=\frac{k-\frac{N}{2}}{\frac{N}{2}}$, we get
\begin{equation} \label{Sk}
	S(k)=S_0-k_B\frac{(k-\frac{N}{2})^2}{\frac{N}{2}}
\end{equation}
where $S_0$ is the entropy of the equilibrium state.
Given that $k$ molecules move in the $+x$ direction, the number of molecules that move in the $-x$ direction is $N-k$. Then the macroscopic $x$-velocity of the whole dilute gas is
\begin{equation}\label{2k} 
	\frac{(2k-N)\bar{v}_z}{N}.
\end{equation}
The same velocity can also be obtained as the average macroscopic directed flow velocity 
\begin{equation} 
	\frac{dv_x}{dz}\frac{\lambda_z}{2}.
\end{equation}
The two expressions are about the same quantity and so they are equal, which leads to
\begin{equation}\label{k} 
	k-\frac{N}{2}=\frac{N\lambda_z}{4\bar{v}_z}\frac{dv_x}{dz}.
\end{equation}
Then by using $N=nA\lambda_z$, we may rewrite Eq. (\ref{Sk}) as
\begin{equation}\label{dilute_gas_entropy}
        S(\frac{dv_x}{dz})=S_0-\frac{k_BnA\lambda_z^3}{8\bar{v}_z^2}\left(\frac{dv_x}{dz}\right)^2.
\end{equation}
Actually, the entropy is still lower, as discussed in the next section, and we write
\begin{equation} \label{dvxdz}
        S(\frac{dv_x}{dz})\approx S_0-\frac{k_BnA\lambda_z^3}{2\bar{v}_z^2}\left(\frac{dv_x}{dz}\right)^2. 
\end{equation}
This is the entropy for a dilute gas that has height of only $\lambda_z$. 

\section{Discussion}
In our study, we have chosen the height of the dilute gas as $\lambda_z$. By choosing height $\lambda_z$, a molecule can collide both with other molecules and the bottom plate, which is stationary. When a molecule collides with the bottom plate, its average velocity remains about zero; when molecules collide with each other, a velocity gradient may appear as a result of a velocity distribution fluctuation.

We have written the entropy of such a dilute gas as Eq. (\ref{dvxdz}), which is Eq. (\ref{dilute_gas_entropy}) plus two corrections. 

\begin{figure}[htbp]
  \begin{center}
	\setlength{\unitlength}{1cm}  
	\centering      
	\begin{picture}(60,5)   
		\put(0,0){\epsfxsize=5.0cm\epsfysize=5.0cm\epsffile{having_gradient_vxz.eps}}
    		   \put(6,0){\epsfxsize=5.0cm\epsfysize=5.0cm\epsffile{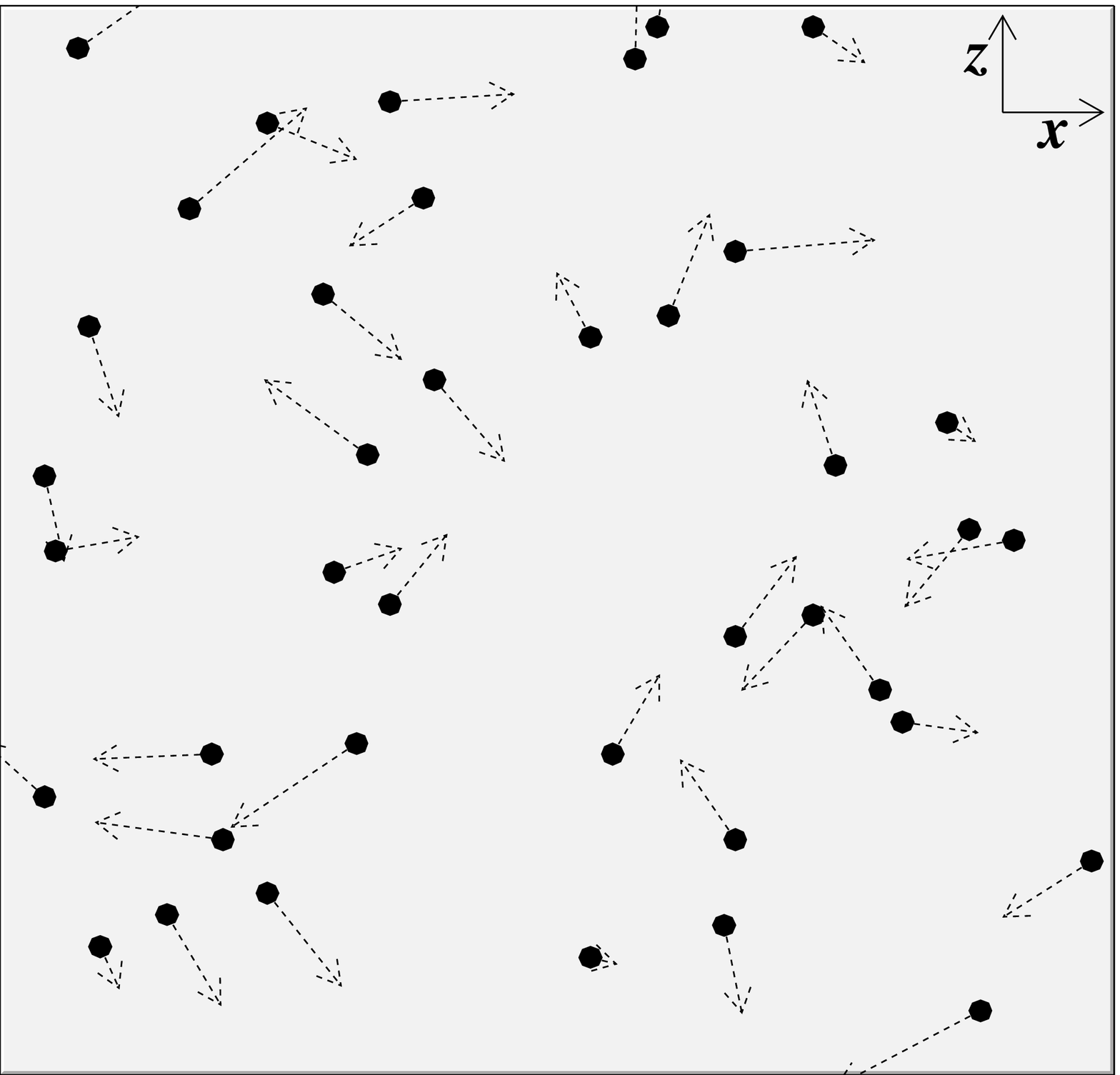}}
    		\put(12,2.6){\epsfxsize=5.0cm\epsfysize=2.5cm\epsffile{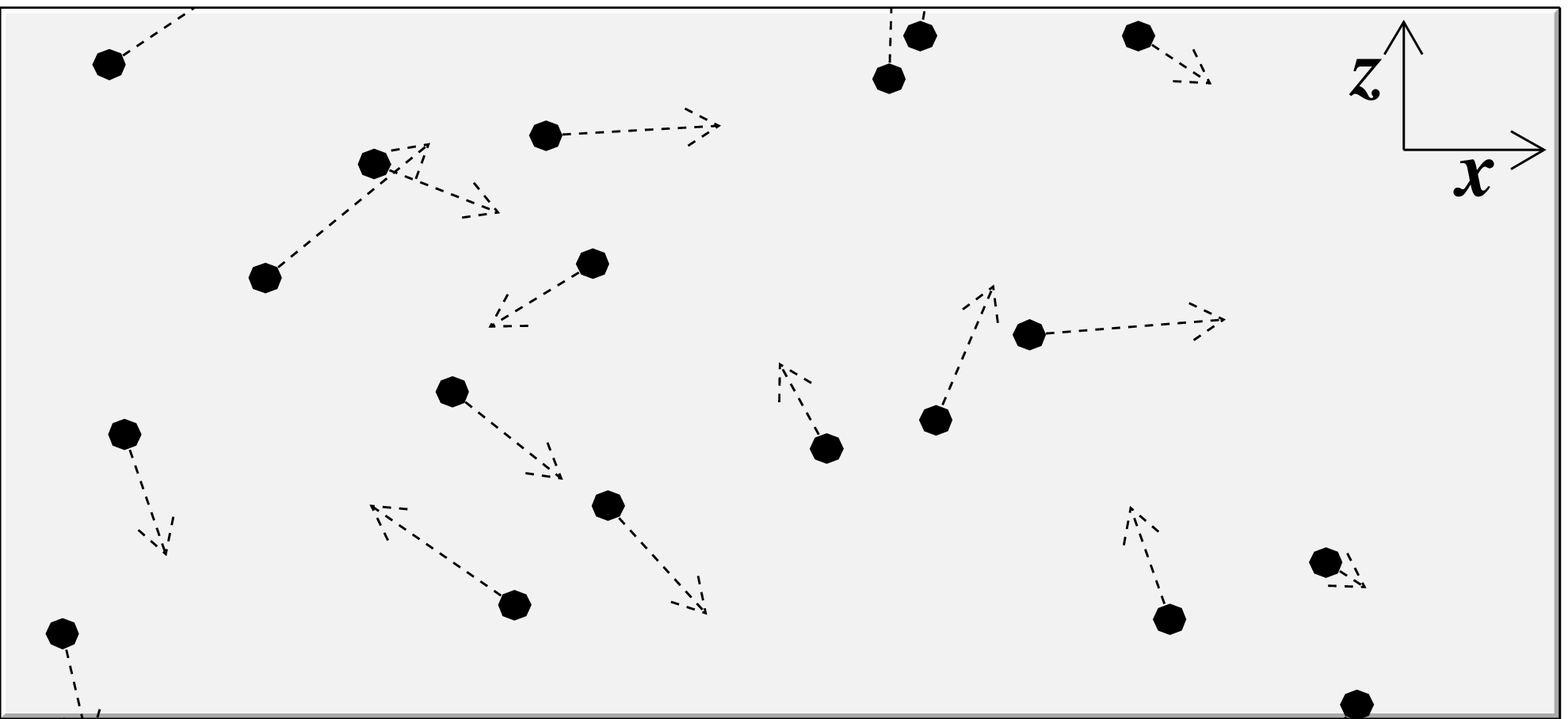}}
		\put(12,0)  {\epsfxsize=5.0cm\epsfysize=2.5cm\epsffile{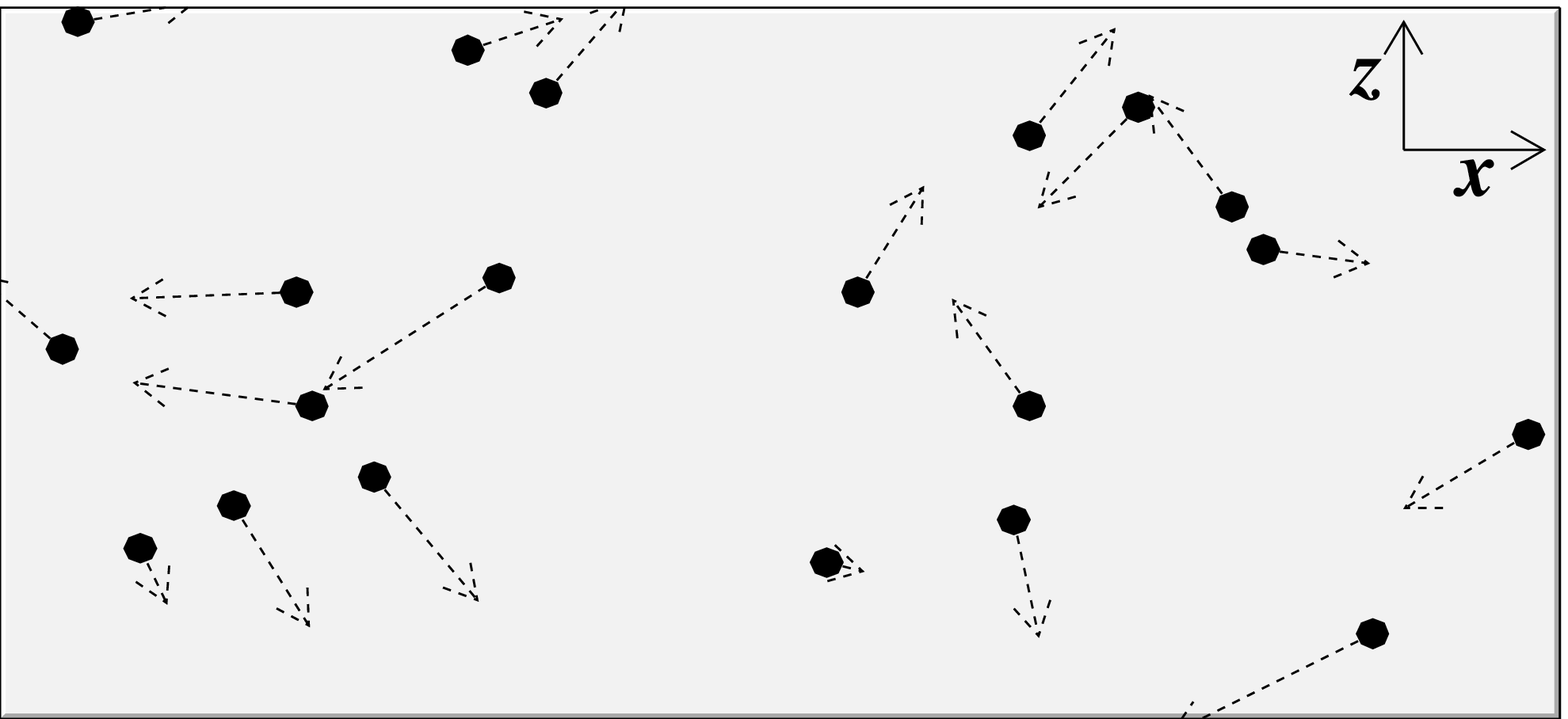}}
	\end{picture}
  \end{center}
\caption{
Additional negative entropy arising due to the uniform velocity gradient distribution. The left graph shows a dilute gas having height $\lambda_z$ and carrying a velocity gradient $\frac{dv_x}{dz}$. The middle graph is for the same dilute gas but has been boosted to the rest frame. The right graph is also for the same dilute gas in the rest frame but has been viewed as two dilute gases each having height $\lambda_z/2$. Then the discussion about Fig. \ref{gradient} applies to each of them, except that the height now becomes $\lambda_z/2$. Thus, additional negative entropy arises and Eq. (\ref{dilute_gas_entropy}) becomes Eq. (\ref{entropy_array}).
  \label{divide}}
\end{figure}

(1) Eq. (\ref{dilute_gas_entropy}) is derived from Eq. (\ref{Sk}). But given a $k$, the corresponding velocity gradient does not necessarily appear, it only appears with an additional feature. The velocity gradient is distributed uniformly across the whole dilute gas, although local fluctuations also exist. Thus the dilute gas can be viewed as two sub-layers where each sub-layer can be analyzed in the same way as the original layer, see Fig. \ref{divide}. By recursively going through this process and neglecting the local fluctuations, Eq. (\ref{dilute_gas_entropy}) becomes 
\begin{align}\label{entropy_array}
\nonumber S(\frac{dv_x}{dz})&=S_0-\frac{k_BnA\lambda_z^3}{8\bar{v}_z^2}\left(\frac{dv_x}{dz}\right)^2-2\frac{k_BnA(\lambda_z/2)^3}{8\bar{v}_z^2}\left(\frac{dv_x}{dz}\right)^2-4\frac{k_BnA(\lambda_z/4)^3}{8\bar{v}_z^2}\left(\frac{dv_x}{dz}\right)^2-\cdots\\
&=S_0-\frac{k_BnA\lambda_z^3}{6\bar{v}_z^2}\left(\frac{dv_x}{dz}\right)^2. 
\end{align}

\begin{figure}[htbp]
  \begin{center}
	\setlength{\unitlength}{1cm}  
	\centering      
	\begin{picture}(60,10)   
		\put(0,0){\epsfxsize=5.0cm\epsfysize=10.0cm\epsffile{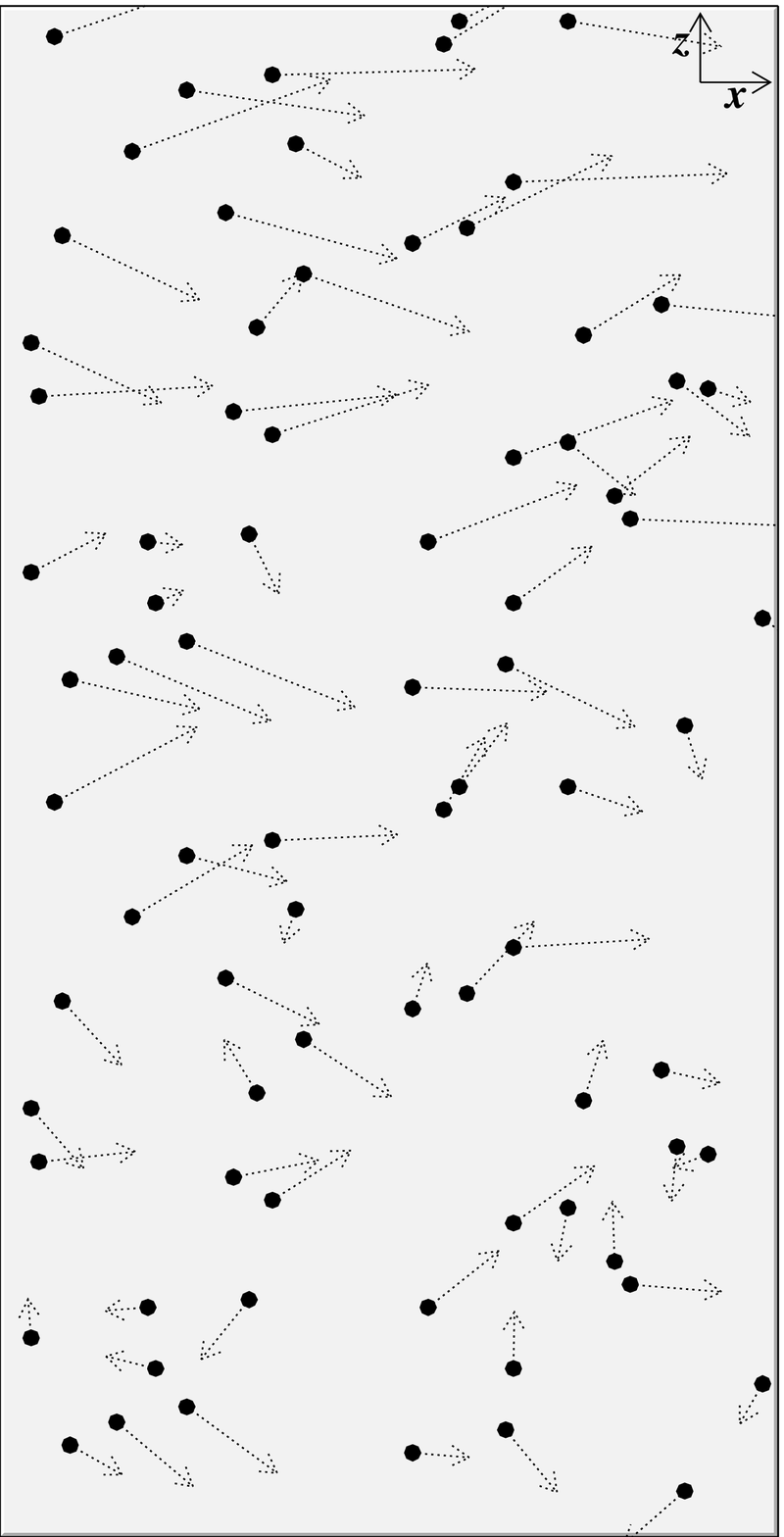}}
		\put(5.5,0){\epsfxsize=5.0cm\epsfysize=10.0cm\epsffile{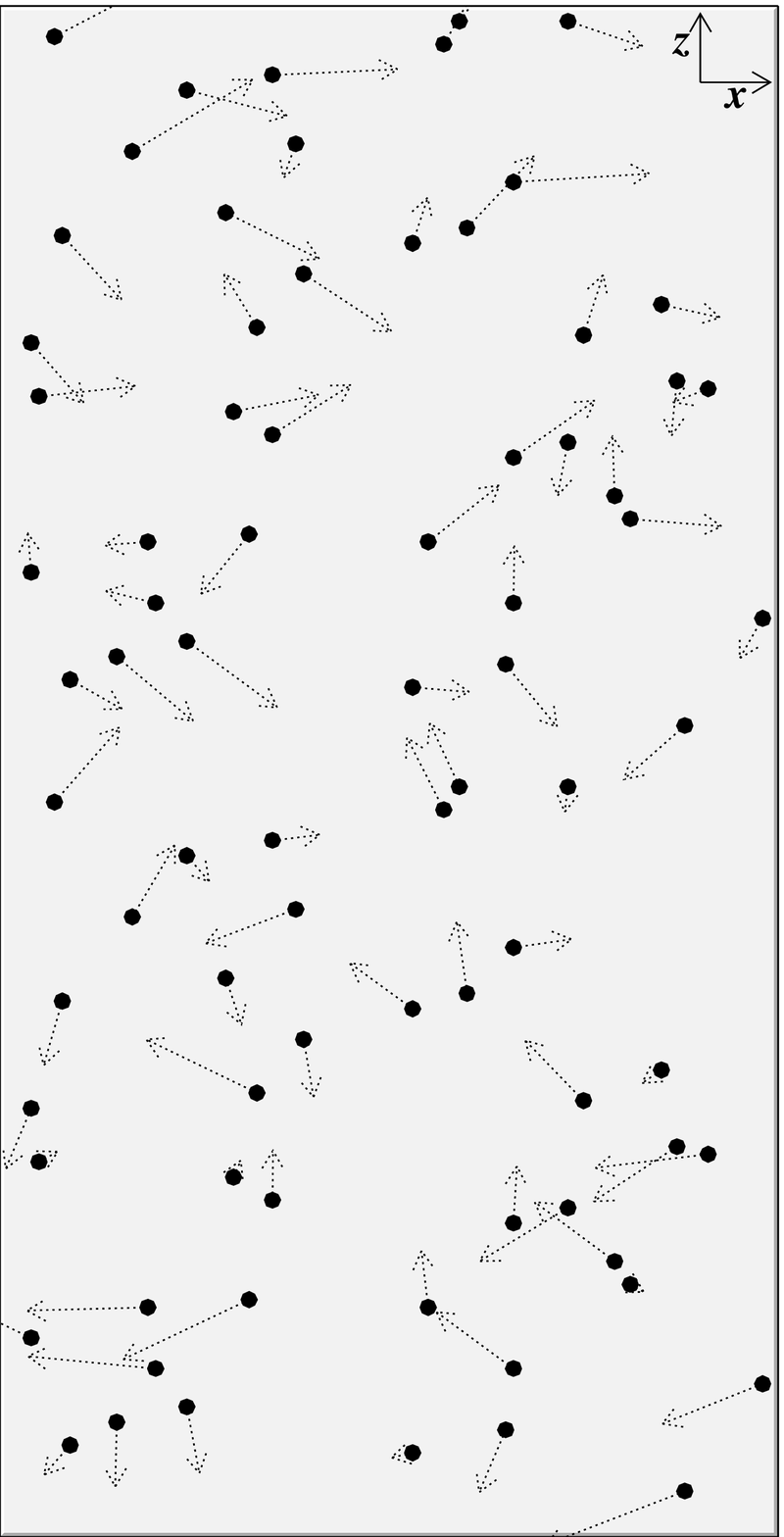}}
		\put(11,7.5){\vector(1,0){1}}
		\put(11,8) {$\frac{dv_x}{dz}\frac{\lambda_z}{2}$}
		\put(12,2.5){\vector(-1,0){1}}
		\put(11,3) {$\frac{dv_x}{dz}\frac{\lambda_z}{2}$}
		\put(12,5.1){\epsfxsize=5.0cm\epsfysize=5.0cm\epsffile{central_frame.eps}}
		\put(12,0){\epsfxsize=5.0cm\epsfysize=5.0cm\epsffile{central_frame.eps}}
	\end{picture}
  \end{center}
\caption{
Additional negative entropy of a dilute gas arising due to the large height. The left graph shows a dilute gas carrying a velocity gradient $\frac{dv_x}{dz}$, its height is $2\lambda_z$. The middle graph is for the same dilute gas but has been boosted to its rest frame. The right graph is for the same dilute gas, but has been viewed as two dilute gases, each of which has height $\lambda_z$ and is in its rest frame, with the two dilute gases moving in opposite directions with velocity $\frac{dv_x}{dz}\frac{\lambda_z}{2}$. For each dilute gas, this velocity leads to a kinetic energy $\frac{\rho A\lambda_z^3}{8}\left(\frac{dv_x}{dz}\right)^2$ and an additional negative entropy $\frac{\rho A\lambda_z^3}{8T}\left(\frac{dv_x}{dz}\right)^2$.
\label{large_scale}}
\end{figure}

(2) Given a velocity gradient $\frac{dv_x}{dz}$, the molecular $x-$speed distribution is not the same as the equilibrium state, even in the center-of-mass frame. Think of a dilute gas that can be viewed as many layers each having height $\lambda_z$; any two adjacent layers will have different macroscopic velocities, see Fig. \ref{large_scale}, and the corresponding kinetic energy is
\begin{equation} 
	2\times\frac{1}{2}\rho A\lambda_z\left(\frac{dv_x}{dz}\frac{\lambda_z}{2}\right)^2=\frac{\rho A\lambda_z^3}{4}\left(\frac{dv_x}{dz}\right)^2.
\end{equation}
This energy, when the velocity gradient relaxes, will become heat and lead to an additional entropy 
\begin{equation} \label{rhoA}
	\frac{\rho A\lambda_z^3}{4T}\left(\frac{dv_x}{dz}\right)^2,
\end{equation}
where $T$ is the temperature. By using $\rho=nM$ and $\bar{v}=\sqrt{\frac{8k_BT}{\pi M}}$, we write this entropy as
\begin{equation}\label{2kBn} 
		\frac{2k_Bn A\lambda_z^3}{\pi \bar{v}^2}\left(\frac{dv_x}{dz}\right)^2=\frac{2k_Bn A\lambda_z^3}{3\pi \bar{v}_z^2}\left(\frac{dv_x}{dz}\right)^2. 
\end{equation}
By this amount, the entropy of the dilute gas is further reduced and so Eq. (\ref{entropy_array}) approximately becomes Eq. (\ref{dvxdz}).


\section{Entropy Production and Viscosity of a Dilute Gas}
In this section, we start with the calculation of the entropy production of a dilute gas. The dilute gas is confined between two plates, with the bottom plate stationary and the top plate subject to a constant shear stress $\tau_{zx}$.
Across the dilute gas, a steady velocity gradient will be established. 
Its expression shall be determined by using the SEA principle in conjunction with the MEP principle. To do that, for a given velocity gradient, we need to examine two specific possibilities of how it evolves and calculate the corresponding entropy production.

One possibility is that the given velocity gradient is steady. For this possibility, the heat generated during unit time interval is 
\begin{equation} 
	\tau_{zx}A\Delta z \frac{dv_x}{dz},
\end{equation}
where $A$ is the cross-sectional area of the dilute gas and $\Delta z$ is the height.
This heat will be absorbed by the dilute gas. The entropy of the dilute gas (or the environment if the heat is transfered to the environment) will increase. The corresponding entropy production is
\begin{equation}\label{steady_entropy_production}
        \sigma=A\Delta z\frac{\tau_{zx}}{T}\frac{dv_x}{dz},
\end{equation}
where $T$ is the temperature of the dilute gas.

The other possibility is that the given velocity gradient $\frac{dv_x}{dz}$ starts to relax at the maximum rate,
\begin{equation}\label{relax}        
\frac{dv_x}{dz}\exp\left(-\frac{t}{\tau}\right),
\end{equation}
where $\tau$ is the minimum possible relaxation time. $\tau$ is approximately $\lambda/\bar{v}$.
To see why the relaxation is exponential and why $\tau\approx \lambda/\bar{v}$, let us study an example of a dilute gas that has height $2\lambda_z$ and is confined between two plates moving in opposite directions at the same speed. A velocity gradient will be established across the dilute gas. Suppose that the two moving plates are suddenly detached from the dilute gas, the velocity gradient will relax gradually at the maximum rate. Let us view the dilute gas as two layers each having height $\lambda_z$, and study their relative velocity. The relative velocity will relax as a result of the momentum exchange between the two layers, but the momentum exchange also depends on the relative velocity and is proportional to it. Thus, the smaller the relative velocity is, the smaller the momentum exchange rate becomes; the smaller the momentum exchange rate becomes, the slower the rate that the relative velocity will relax. As a result, they both decrease exponentially. Thus the velocity gradient relaxes exponentially too. 
 Concerning $\tau$, a molecule needs to take time $\lambda/\bar{v}$ to transfer a momentum from one layer to another, and the mean value of the momentum that it transfers is comparable to the mean molecular momentum difference between the two layers. Thus, within a few transfers for each molecule, the velocity gradient will relax. So the relaxation time $\tau$ is approximately $\lambda/\bar{v}$, but a little bit larger. What we have studied here is a dilute gas that has height $2\lambda_z$. For a dilute gas that is thicker, it can be viewed as many layers that each has height $\lambda_z$. Then, between any two adjacent layers, the discussion is the same. But each layer except for the two end layers has two adjacent layers and thus all layers are correlated. This makes the relaxation time larger. But we are only concerned with the instant at which the velocity gradient starts to relax; at that instant, the relaxation correlation has not yet taken effect, and thus, in the limit $t\to 0$, $\tau$ should be the same for a dilute gas having any height.

Given the velocity gradient Eq. (\ref{relax}), Eq. (\ref{dvxdz}) leads to 
\begin{equation}
	S(t)=S_{0}-\frac{k_BnA\lambda_z^3}{2\bar{v}_z^2}\left(\frac{dv_x}{dz}\right)^2\exp\left(-2\frac{t}{\tau}\right),
\end{equation}
and the corresponding entropy production is
\begin{equation}
	\sigma(t)=\frac{dS(t)}{dt}=\frac{k_BnA\lambda_z^3}{\tau \bar{v}_z^2}\left(\frac{dv_x}{dz}\right)^2\exp\left(-2\frac{t}{\tau}\right),
\end{equation}
which, in the limit $t\to 0$, leads to
\begin{equation}\label{sigma_S}
	\sigma=\frac{k_BnA\lambda_z^3}{\tau \bar{v}_z^2}\left(\frac{dv_x}{dz}\right)^2.
\end{equation}
This is the entropy production of a dilute gas that has height $\lambda_z$. For a dilute gas thicker, in the limit $t\to 0$, it can be viewed as many such layers and the entropy production is then 
\begin{equation} \label{relax_entropy_production}
        \sigma=\frac{\Delta z}{\lambda_z}\frac{k_BnA\lambda_z^3}{\tau \bar{v}_z^2}\left(\frac{dv_x}{dz}\right)^2=A\Delta z\frac{k_Bn\lambda^2}{\tau \bar{v}^2}\left(\frac{dv_x}{dz}\right)^2.
\end{equation}
This is because at the instant when the velocity gradient starts to relax, any two non-adjacent layers should have no relaxation correlation established yet. The reason why we use the limit $t\to 0$ is that we only need to know whether or not the given velocity gradient would start to relax. The relaxation process itself is irrelevant.

\begin{figure}[htbp]
  \begin{center}
    \mbox{\epsfxsize=12.0cm\epsfysize=10.0cm\epsffile{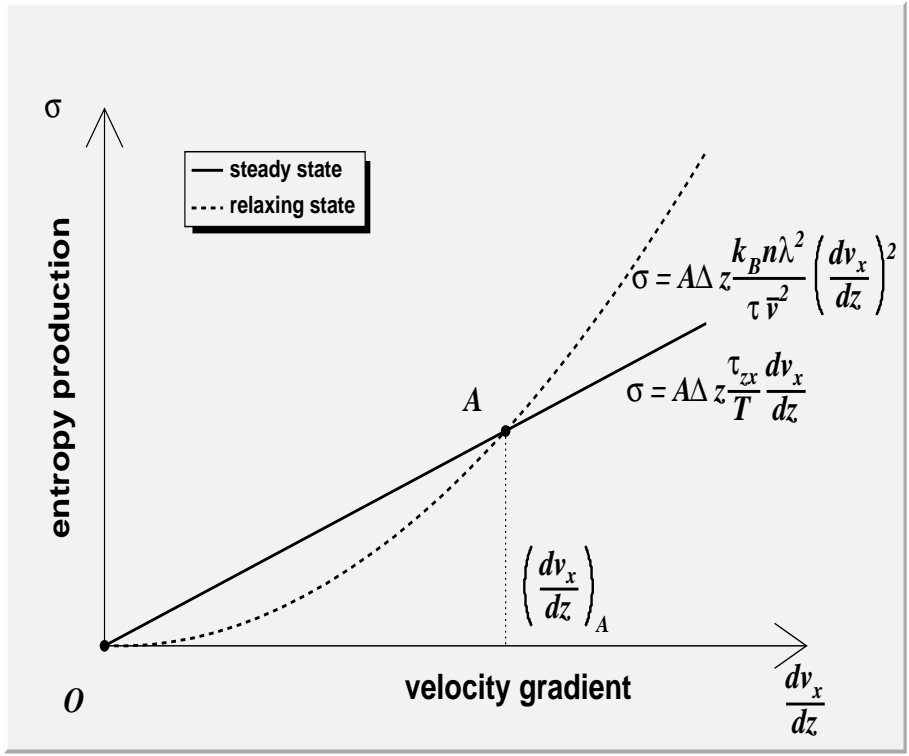}}
  \end{center}
\caption{Entropy production with respect to small velocity gradient. The solid curve is the entropy production when the velocity gradient is steady. The dashed curve is the entropy production when the velocity gradient starts to relax at the maximum rate. $(\frac{dv_x}{dz})_A$, the velocity gradient at point $A$, is the determined steady velocity gradient. For a velocity gradient $\frac{dv_x}{dz}>(\frac{dv_x}{dz})_A$, it cannot be steady, because in this range the solid curve is not the highest, at least it is lower than the dashed curve. The steepest entropy ascent (SEA) principle indicates that a state can occur only if it is associated with the maximum entropy production. Thus the steady velocity gradient must be in the range $0\le \frac{dv_x}{dz}\le (\frac{dv_x}{dz})_A$. Subsequently, according to the maximum entropy production (MEP) principle, we know that the steady velocity gradient must be $(\frac{dv_x}{dz})_A$, because among all candidates it is associated with the maximum entropy production.\label{fig1}}
\end{figure}

The entropy productions of the two possibilities are plotted in Fig. \ref{fig1}. They are equal when $\frac{dv_x}{dz}=\left(\frac{dv_x}{dz}\right)_A$. We shall use first the SEA principle and then the MEP principle to determine that the steady velocity gradient must be $\left(\frac{dv_x}{dz}\right)_A$. According to the SEA principle, for a velocity gradient $\frac{dv_x}{dz}>\left(\frac{dv_x}{dz}\right)_A$, it cannot be steady because the entropy production would not be the maximum. (But it would not start to relax at the maximum rate either; actually, the velocity gradient will start to relax at a smaller rate until $\frac{dv_x}{dz}=\left(\frac{dv_x}{dz}\right)_A$, and during the entire relaxation process the shear stress will continue to generate heat and entropy.) Thus the steady velocity gradient must be in the range $0\le \frac{dv_x}{dz}\le \left(\frac{dv_x}{dz}\right)_A$. Then according to the MEP principle, we see that the steady velocity gradient must be $\left(\frac{dv_x}{dz}\right)_A$ because, among all candidates, it is the one that is associated with the maximum entropy production.

Another way to think about Fig. \ref{fig1} is that the two entropy productions compete with each other. 
Given a dilute gas filled between a stationary bottom plate and a moving upper plate, the dilute gas itself tends to relax toward the equilibrium state, while the upper moving plate tends to move faster and faster given the shear stress being constant. But the two tendencies are in opposite directions and compete with each other, until at last one balances the other and the velocity gradient becomes steady. Even so, the relaxing tendency still presents. And its strength can be measured by the entropy production shown in Eq. (\ref{relax_entropy_production}). So, for a given velocity gradient to be steady, it must at least overcome that relaxing tendency by having an equal or larger entropy production. Thus, its upper limit is obtained from Fig. \ref{fig1} as $\left(\frac{dv_x}{dz}\right)_A$. At the same time, the upper plate still tends to move faster, and its strength should be measured by another entropy production, which must be overcome too if a velocity gradient is to be steady. Thus the lower limit of the steady velocity gradient should also exist. And perhaps the upper limit equals the lower limit and they are the steady velocity gradient. If so, the steady velocity gradient can be obtained on the basis of the SEA principle alone.

By calculating $\left(\frac{dv_x}{dz}\right)_A$ in Fig. \ref{fig1}, the steady velocity gradient is obtained as
\begin{equation}\label{j}
        \frac{dv_x}{dz}=\frac{\tau\bar{v}^2}{k_BTn\lambda^2}\tau_{zx}.
\end{equation}
So we have
\begin{equation} 
	\tau_{zx}=\frac{k_BTn\lambda^2}{\tau\bar{v}^2}\frac{dv_x}{dz}.
\end{equation}
Comparing it with Eq. (\ref{jv}), the viscosity of the dilute gas is extracted as
\begin{equation}\label{kappa2}
         \eta=\frac{k_BTn\lambda^2}{\tau\bar{v}^2}.
\end{equation}
By using relations  
\begin{equation} 
	\bar{v}=\sqrt{\frac{8k_BT}{\pi M}}, \ \rho = nM, \ \tau\sim \frac{\lambda}{\bar{v}},
\end{equation}
we may write it as
\begin{equation} 
	\eta=\frac{\pi}{8}\rho \bar{v} \lambda,
\end{equation}
which is comparable to the kinetic theory result (\ref{kd}).

\section{Conclusion}
We have derived the viscosity of a dilute gas by using a new approach. The new approach is based on the fact that there exists an entropy production competition between at least two possibilities of how a given velocity gradient evolves. One possibility is that the velocity gradient is steady. The other possibility is that the velocity gradient relaxes exponentially at the maximum rate. Then, by applying the steepest entropy ascent (SEA) principle, we have obtained the upper limit of the steady velocity gradient. Subsequently, by applying the maximum entropy production (MEP) principle, we have identified the actual steady velocity gradient from whose expression the viscosity is extracted. The viscosity extracted is comparable to the result of kinetic theory.

{\bf Acknowledgment:} Work supported by Liaoning Education Office Scientific Research Project(2008288), and by SRF for ROCS, SEM.

\end{document}